\documentclass[journal=jacsat,manuscript=article]{achemso}   
\usepackage{amsfonts}
\usepackage{amssymb}
\usepackage{graphicx}
\usepackage{amsmath}
\usepackage{tikz}
\usepackage{adjustbox}
\usepackage[normalem]{ulem}
\usetikzlibrary{shapes.geometric, arrows}
\usepackage[english]{babel}
\usepackage{color}
\usepackage[version=3]{mhchem} 
\usepackage{natbib}
\usepackage{url}
\usepackage[colorlinks,citecolor=black,urlcolor=blue,linkcolor=black]{hyperref} 

\usetikzlibrary{arrows,shapes,positioning,shadows,trees}

\tikzstyle{bag} = [align=center]

\newcommand{\olr}[1]{{\color{red}{}}}

\newcommand{\osvp}[1]{{\color{blue}{}}}

\author{Sebastian V. Pios}
\affiliation{Zhejiang Laboratory, Hangzhou 311100, China}
\author{Maxim F. Gelin}
\affiliation{School of Science, Hangzhou Dianzi University, Hangzhou 310018, China}
\author{Luis Vasquez}
\affiliation{School of Science, Hangzhou Dianzi University, Hangzhou 310018, China}
\author{J{\"u}rgen Hauer}
\affiliation{Department of Chemistry, Technical University of Munich, D-85747 Garching, Germany}
\author{Lipeng Chen}
\affiliation{Zhejiang Laboratory, Hangzhou 311100, China}
\email{chenlp@zhejianglab.com}

\title{On-the-Fly Simulation of Two-Dimensional Fluorescence-Excitation Spectra}

\begin{document}


\begin{abstract}
Two-dimensional (2D) fluorescence-excitation (2D-FLEX) spectroscopy is a recently proposed nonlinear femtosecond technique for the detection of photoinduced dynamics. The method records a time-resolved fluorescence signal in its excitation- and detection-frequency dependence, and hence combines the exclusive detection of excited state dynamics (fluorescence) with signals resolved in both excitation and emission frequencies (2D electronic spectroscopy). In this work, we develop an on-the-fly protocol for the simulation of 2D-FLEX spectra of molecular systems, which is based on interfacing the classical doorway-window representation of spectroscopic responses with trajectory surface hopping simulations. Applying this methodology to the gas-phase pyrazine, we show that femtosecond 2D-FLEX spectra can deliver detailed information otherwise obtainable via attosecond spectroscopy. 
\end{abstract}



Two-dimensional electronic spectroscopy (2D-ES) was developed at the turn of the millennia as the optical analogue to nuclear magnetic resonance (NMR) spectroscopy. \cite{Jonas03,Warren03,Hochstrasser07} Since then, 2D-ES has become one of the most powerful and information-rich femtosecond four-wave-mixing techniques in the UV/Vis regimes. It  was implemented in collinear, \cite{Wagner05} partially non-collinear,\cite{DeFlores07,Riedle13,Ogilive15} and fully non-collinear realizations \cite{Graham04, Milota13} and emerged as an efficient high-precision technique for the comprehensive real-time characterization of a large variety of material systems, from atoms \cite{Stienkemeier21} and polyatomic chromophores\cite{Chergui21} to molecular aggregates and solids. \cite{Brixner05,engel2007evidence,Lim2015,Collini21,Scholes22}

\begin{figure}
\includegraphics[scale=0.5]{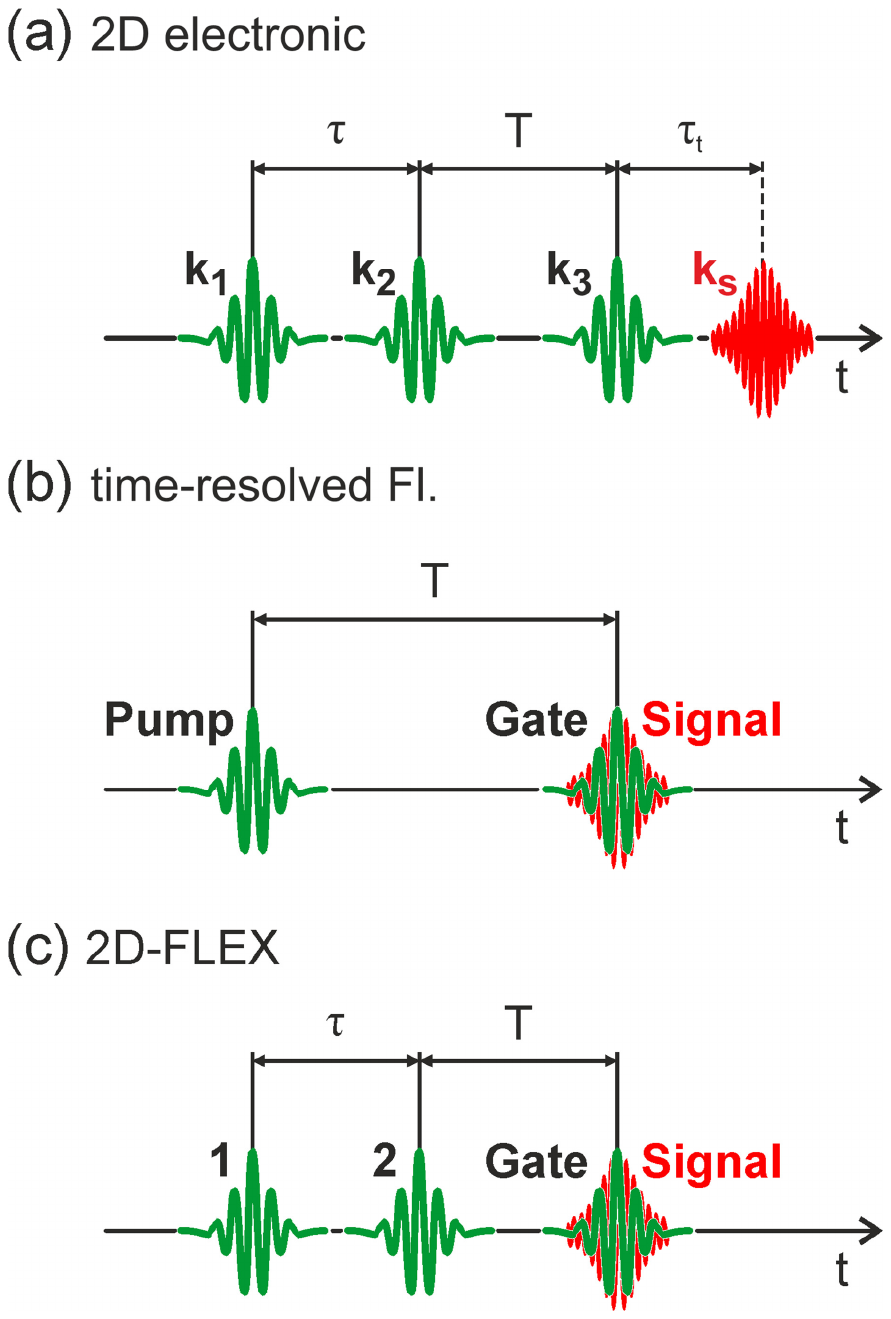}
\caption{Pulse sequences shaping 2D-ES (a), time- and frequency-resolved fluorescence (b), and 2D-FLEX (c) signals.
\label{Pulses}}
\end{figure}

 2D-ES is a variant of the third-order heterodyne-detected four-wave-mixing spectroscopy (see Figure \ref{Pulses}(a)). The 2D-ES signal $S^{(\sigma)}(\tau,T,\tau_{t})$ can be recorded in the rephasing ($\sigma=R$, $\mathbf{k_{S}}=\mathbf{k_{1}}-\mathbf{k_{2}}+\mathbf{k_{3}}$) or non-rephasing ($\sigma=NR$, $\mathbf{k_{S}}=-\mathbf{k_{1}}+\mathbf{k_{2}}+\mathbf{k_{3}}$) phase-matching direction as a function of the delay times between the first two pulses (coherence time $\tau$), the second and the third pulses (population time $T$), and the last two pulses (detection time $\tau_{t}$). The sketch of Figure~\ref{Pulses}(a) indicates that 2D-ES delivers four-wave-mixing signal with maximal information content as both excitation and emission frequencies are fully resolved. 
 This allows for an unprecedented level of scrutiny of material systems, explaining why 2D-ES is so powerful and popular. 2D-ES signals  are usually Fourier transformed with respect to $\tau$ (excitation frequency $\omega_\tau$) and $\tau_t$ (detection frequency $\omega_t$), and the resultant 2D-ES spectra $S^{(\sigma)}(\omega_{\tau}, T, \omega_t)$ visualize how correlation between the initial  ($\omega_{\tau}$) and final ($\omega_{t}$) states of the system evolves with the population time $T$.

2D-ES spectra consist of three contributions, which are customarily referred to as  ground-state bleach (GSB), stimulated emission (SE) and excited state absorption (ESA) 
\begin{equation}
		S^{(\sigma)}(\omega_{\tau},T,\omega_{t})=
  \sum_{j=\text{GSB, SE, ESA}}
  S_{j}^{(\sigma)}(\omega_{\tau},T,\omega_{t}).\label{2Di}
	\end{equation}
However, the information content of these contributions differs substantially. GSB monitors wavepacket motion on the electronic ground state, while SE and ESA reflect wavepacket dynamics on the excited electronic states. Yet, SE and ESA differ in their respective ``spectator states" which are responsible for encoding the dynamical information. In the SE, the wavepacket motion on the excited electronic states is projected onto the electronic ground state, while it is projected onto higher-lying excited electronic states in the ESA. For complex material systems, it is difficult to decipher the intrinsic system dynamics from GSB, SE, and notably ESA spectra. Furthermore, GSB, SE and ESA spectra usually overlap, which leads to the partial cancellation of spectral features and hinders proper interpretation of 2D-ES spectra. While the techniques of  polarization-sensitive detection  \cite{Hochstrasser01,Gelin13a,Zanni22} or beating maps \cite{Leonas14,EET1,EET3} can partially solve the problem, they do not provide a universal solution. 

As has recently been proposed by Hauer and coworkers,\cite{2d_flex_2023} this problem can be addressed by a novel technique,  2D fluorescence-excitation (2D-FLEX) spectroscopy, which combines the spectral resolution of excitation and detection processes inherent in 2D-ES with the exclusive detection of the SE contribution. The principle of 2D-FLEX can readily be grasped from the following considerations. Time-resolved fluorescence spectroscopy \cite{chem_rev_vauthey2017} permits a direct detection of wavepacket dynamics on the excited electronic states, but cannot provide information on electronic coherences during the excitation of the system with a single pump pulse (see Figure \ref{Pulses}(b)). In 2D-FLEX, a single pump pulse is replaced by a pump-pulse pair delayed by a time $\tau$, as sketched  in Figure \ref{Pulses}(c). Hence, 2D-FLEX can be envisaged as a technique  combining the double-pulse excitation of 2D-ES with the broadband detection of time- and frequency-resolved fluorescence. 

The high information content of 2D-FLEX spectroscopy has been demonstrated by simulating spectra for several model systems.\cite{2d_flex_2023} Yet, there is still a lack of practical tools to simulate 2D-FLEX spectra for realistic molecular systems. The aim of this work is to develop an \textit{ab initio} theoretical framework for the simulation of 2D-FLEX spectra of molecular systems by interfacing the classical doorway-window representation of spectroscopic responses with trajectory surface hopping simulations. The method has been implemented for the pyrazine molecule, a system known for its multiple conical intersections of lower lying excited electronic states. The photophysics of pyrazine has been comprehensively studied by \textit{ab initio} quantum-chemistry methods,\cite{woywod_pyrazine,gatti_pyrazine} dynamic calculations,\cite{xie_pyrazine,CLP} and simulations of nonlinear spectroscopic signals. \cite{OTFDW1,Xiang2D,Skw20a,LP21,OTFDW2,OTFDW4,pios_pyrazine2024} 

The 2D-FLEX spectra (see Figure~\ref{Pulses}(c)) can be represented as a sum of rephasing and nonrephasing contributions,
\begin{equation}\label{2DF}
S_{2DF}(\omega_{\tau},T,\omega_{t}) \sim \mathrm{Re} 
\sum_{\sigma=R, \, NR} S_{2DF}^{(\sigma)}(\omega_{\tau},T,\omega_{t}),    
\end{equation}
which can be further evaluated in terms of third-order response functions as follows \cite{2d_flex_2023}
	\begin{equation}
		S_{2DF}^{(\sigma)}(\omega_{\tau},T,\omega_{t}) \sim \, \int_{0}^{\infty}d\tau\, 
    \int_{-\infty}^{\infty}dt\,\int_{0}^{\infty}dt_{3}\,\int_{0}^{\infty}dt_{2}\,\int_{0}^{\infty}dt_{1}\,R_{\sigma}(t_{3},t_{2},t_{1})e^{i\eta_{\sigma}\omega_{\tau}\tau}\times\label{Inr}
	\end{equation}
	\[
	 e^{i\eta_{\sigma}\omega_{pu}(t_{1}-\tau)}e^{-(\zeta-i\omega_{t})t_{3}}E_{t}(t)E_{t}(t-t_{3})E_{pu}(t+T-t_{3}-t_{2})E_{pu}(t+\tau+T-t_{3}-t_{2}-t_{1}).
	\]
Here $\sigma=R,NR$, $\eta_{NR}=1$, $\eta_{R}=-1$; $\omega_{pu}$  and $E_{pu}(t)$ represent the carrier frequency and the dimensionless envelope of the pump pulse;  $\omega_{t}$ is the fluorescence frequency, and $E_t(t)$ is the envelope of the up-conversion (gate) pulse which is responsible for the time resolution of the emitted fluorescence; $t_{1}$, $t_{2}$, and $t_{3}$ are the time intervals between sequential system-field interactions;  $R_{NR}(t_{3},t_{2},t_{1})$ and $R_{R}(t_{3},t_{2},t_{1})$ are the non-rephasing and rephasing  response functions which correspond, respectively, to the response functions $R_{1}(t_{3},t_{2},t_{1})$ and $R_{2}(t_{3},t_{2},t_{1})$ of Mukamel's monograph.\cite{MukamelBook}

2D-FLEX spectra $S_{2DF}(\omega_{\tau},T,\omega_{t})$ of Eq. (\ref{2DF}) and  SE contributions to the total (rephasing + nonrephasing) 2D-ES spectra 
\begin{equation}\label{SE}
S_{SE}(\omega_{\tau},T,\omega_{t}) \sim \mathrm{Re} 
\sum_{\sigma=R, \, NR} S_{SE}^{(\sigma)}(\omega_{\tau},T,\omega_{t}),    
\end{equation}
are  determined by the same response functions and are therefore quite similar. Yet, there is a fundamental difference between these two spectra.\cite{2d_flex_2023} The time ($T$) -- frequency ($\omega_{t}$) resolution of 2D-FLEX spectra is Fourier-limited ($\delta_{T}\delta_{\omega_{t}}\sim 1$) because it is governed by the same gate pulse (Figure~\ref{Pulses}(c)), while the time-frequency resolution of 2D-ES spectra is not Fourier limited because it is determined by time delays between different pairs of pulses (Figure~\ref{Pulses}(a)). This observation has profound effect on the peak shapes and time-frequency resolution of 2D-FLEX and 2D-ES SE spectra. While it is instructive to compare 2D-FLEX and 2D-ES SE spectra, we note that only 2D-FLEX spectra can be measured directly. There is no experimentally feasible way to measure the SE-part of a 2D-ES spectrum.

Interfacing  dynamical trajectory methods with \textit{ab initio} electronic-structure
methods into on-the-fly simulation protocols is a flourishing field of research in theoretical spectroscopy (see recent
reviews \cite{Vanicek17,Garavelli18,Garavelli20,Jansen21,Isborn21,Santoro21,Maxim2002CR,Cocchi24}). In the context of the present work, we require a simulation protocol that can account for finite pulse durations, which, as explained above, determine the time-frequency resolution of 2D-FLEX spectra. The protocol which is based on combining trajectory surface hopping (SH) simulations
\cite{Tully,OTF2} with the classical doorway-window (DW)
representation of four-wave-mixing signals excellently meets this criterion.\cite{Maxim2002CR} This protocol was developed and applied for the evaluation of transient-absorption pump-probe,\cite{OTFDW1} 2D-ES, \cite{Xiang2D} and time-resolved fluorescence spectra.\cite{ZL24} Its validity to predict pump-probe and 2D-ES spectra of pyrazine has recently been proved by comparing with full quantum simulations of the models of reduced dimensionality. \cite{pios_pyrazine2024}

Since 2D-FLEX spectra combine   a double-pulse excitation of 2D-ES with the detection of time- and frequency-resolved fluorescence, the DW-SH simulation protocol for 2D-FLEX spectra can be immediately constructed by combining the doorway function of 2D-ES spectra \cite{Xiang2D} with the window function of time-resolved fluorescence spectra.\cite{ZL24} The  computational details are presented in Supporting Information, while the flowchart clarifying the simulation procedure is depicted in Figure~\ref{fig:flowchart}. In the DW-SH simulation protocol, 2D-FLEX spectra can be evaluated as a product of the doorway and window functions averaged over SH trajectories 
\begin{equation}\label{2DF1}
S_{2DF}(\omega_{\tau},T,\omega_{t}) \sim     
\langle D_{2DF}(\omega_{\tau},\boldsymbol{R}_{g},\boldsymbol{P}_{g};e)W_{2DF}(\omega_{t},\boldsymbol{R}_{e}(T),\boldsymbol{P}_{e}(T); e(T))\rangle.
\end{equation}
Here, $\boldsymbol{R}_{g}$ and $\boldsymbol{P}_{g}$ are initial nuclear coordinates and momenta, which can be sampled from the Wigner distribution in the electronic ground state, $\rho_{g}^{\mathrm{Wig}}(\boldsymbol{R}_{g},\boldsymbol{P}_{g})$; \cite{Wigner84} $D_{2DF}(\omega_{\tau},\boldsymbol{R}_{g},\boldsymbol{P}_{g};e)$ is the doorway function of Eq. (S3) in the initial electronic state $e$ sampled according to its oscillator strength and pump-frequency detuning; \cite{Maxim2002CR} 
$W_{2DF}(\omega_{t},\boldsymbol{R}_{e}(T),\boldsymbol{P}_{e}(T); e(T))$ is the window function of Eq. (S4) in which  $\boldsymbol{R}_{e}(T), \, \boldsymbol{P}_{e}(T)$ denote nuclear positions and momenta in the electronic state $e(T)$ at the time  $T$, and the notation $e(T)$ emphasizes that a trajectory initiated at \textit{T}=0~fs in an excited state $e$ may end up at time $T$ in another excited state $e(T)$ due to nonadiabatic transitions; $\langle...\rangle$ denotes averaging over SH trajectories.  The doorway-window expressions for the evaluation of 2D-ES SE spectra \cite{Xiang2D} are also presented in Supporting Information.  
It should be noted that the doorway and window functions are fully specified by electronic energies and transition dipole moments along SH trajectories, as well as by the carrier frequencies and envelops of the pump and gate pulses.  

In all simulations of the present work, 
the pump pulse is tuned into resonance with the bright state ($B_{2u}$) of pyrazine ($\omega_{p}=5.2$ eV), and 
the envelopes of the pump and gate (for 2D-FLEX) and probe (for 2D-ES SE) pulses are modeled by Gaussian functions, $E_a(t)=\exp(-(t/\tau_a)^2)$, $E_a(\omega)=\exp(-(\omega\tau_a)^2/4)$, where $\tau_a$ is the pulse duration ($a=pu, \, t, \, pr$ correspond to pump, gate and probe pulse, respectively). 
The widths of the spectral features of 2D-FLEX and 2D-ES SE spectra are determined by the spectral widths of the laser pulses and  by the peak shape broadening (see Eq. (S5)), in which the inhomogeneous broadening  parameter is fixed at $\tau_{\nu}^{-1}=0.05$~eV in all our calculations.

\begin{figure}
    \centering
    \includegraphics[width=.6\textwidth]{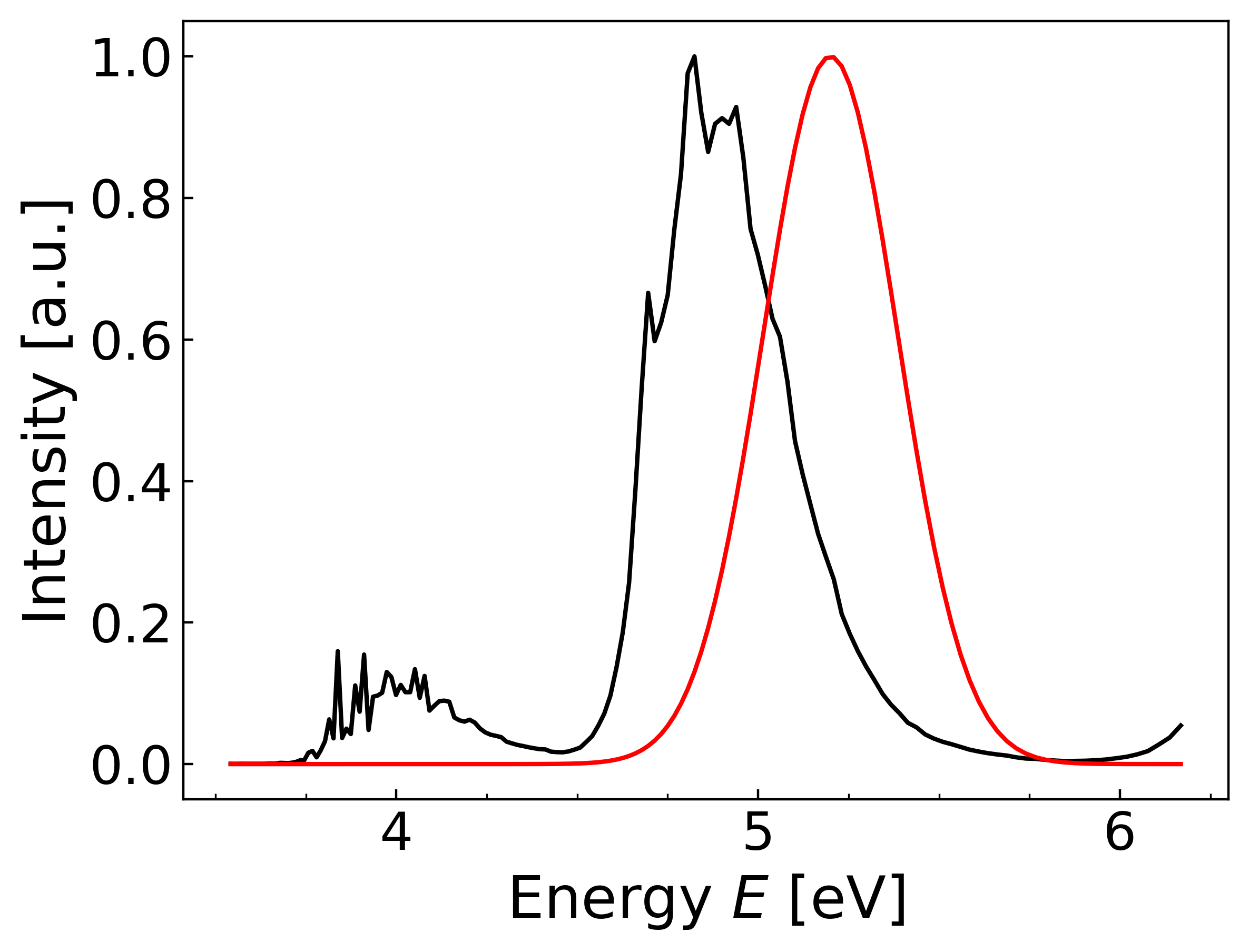}
    \caption{Black: experimental absorption spectrum  of pyrazine molecule in the gas-phase at 353$\pm$3~K (data is taken from Ref.~\cite{Samir2020}); Red: energy profile of the Gaussian laser pulse centered at 5.2 eV with width of $\tau =$ 5~fs.}
    \label{pulses}
\end{figure}

Before we delve into discussion of the simulated 2D-FLEX and 2D-ES SE spectra, it is instructive to briefly characterize the main photophysical processes in pyrazine and determine their spectroscopic manifestations. Pyrazine possesses three low-lying  excited electronic states in the Franck-Condon region, the $B_{2u}(\pi\pi^{*})$, $B_{3u}(n\pi^{*})$ and $A_{u}(n\pi^{*})$, which are nonadiabatically coupled through a series of conical  intersections.\cite{woywod_pyrazine,gatti_pyrazine} The experimental gas-phase UV linear absorption spectrum of pyrazine at 353$\pm$3~K is shown in Figure \ref{pulses} (black, data is taken from Ref. \cite{Samir2020}). The spectrum consists of two bands, an intense band centered around 4.9~eV and a weaker band centered around 4.0~eV. These bands are attributed to the bright $B_{2u}$ state and the weakly absorbing $B_{3u}/A_u$ states, respectively. It is essential that the shape and intensity of the $B_{3u}/A_u$ band is mostly determined by the nonadiabatic coupling between the bright $B_{2u}$ state and almost dark  $B_{3u}/A_u$ states (the so-called intensity borrowing effect). As revealed by the femtosecond photoelectron spectroscopy measurements, the $B_{2u}-B_{3u}$ coupling results in the $\sim20$ fs conical-intersection-driven $B_{2u}\rightarrow B_{3u}$ population transfer.\cite{radlof_pyr_phelsp,suzuki2010time,suzuki2016full} Internal conversion from the coupled $B_{2u}/B_{3u}$ states to the electronic ground state occurs on a much longer timescale of $\sim20$~ps.\cite{radlof_pyr_phelsp,suzuki2010time,suzuki2016full} Hence, only $B_{2u}/B_{3u}$ steady-state fluorescence of pyrazine in the gas phase was measured \cite{Pyrazine_Adv} and simulated.\cite{Domcke90,Domcke92} On the other hand, the previous simulations of time-resolved fluorescence of pyrazine in various reduced-dimensional models \cite{Skw20a,LP21,LP_SE} as well as recent \textit{ab initio} simulations of the SE contribution to pyrazine's transient-absorption pump-probe spectra \cite{OTFDW1,pios_pyrazine2024} reveal that the  $B_{2u}/B_{3u}$ fluorescence of pyrazine survives on a timescale of at least 150 fs, but experiences a fivefold intensity reduction caused by the $B_{2u}/B_{3u}$ conical intersection.

\definecolor{white}{RGB}{255, 255, 255}
\definecolor{black}{RGB}{0, 0, 0}
\definecolor{grey}{RGB}{51, 51, 51}
\tikzset{
    rectanglestyle_round/.style={rectangle, rounded corners, minimum width=3cm, minimum height=1cm, text centered, font=\normalsize, color=black, draw=red, line width=1, fill=white},
    rectanglestyle/.style={rectangle, minimum width=2cm, minimum height=1cm, text centered,, font=\normalsize, color=black, draw=red, line width=1, fill=white},
    diamondstyle/.style={diamond, aspect=2.5, inner xsep=1mm, minimum width=1cm, minimum height=1cm, text width=4cm, text centered, font=\normalsize, color=black, draw=red, line width=1, top color=white, bottom color=blue!30},
    decision/.style={diamond, text centered, draw=black, aspect=5, inner xsep=1mm, top color=white, bottom color=blue!30},
    stop/.style={rectangle, rounded corners, minimum width=3cm, minimum height=1cm,text centered, draw=black},
    arrow_style/.style = {thick, draw=black, line width=2, ->, >=stealth}
    }
\begin{figure}[!t]
    \centering
    \hspace{0cm}
    \begin{adjustbox}{width=0.7\textwidth}
    \begin{tikzpicture}[node distance=3cm]
\node (initialization)[rectanglestyle, bag]{Sample the initial $\boldsymbol{R}_{g}$,  $\boldsymbol{P}_{g}$, $e$  and \\  evaluate the doorway function $D_{2DF}(\omega_{\tau},\boldsymbol{R}_{g},\boldsymbol{P}_{g};e)$};
\node (propagation)[below = 1cm and 3cm of initialization, rectanglestyle, bag] {Propagate and store the stochastically selected classical\\ trajectories using a surface-hopping algorithm\\ for a selected simulation time};
\node (evaluation) [below = 1cm and 3cm of propagation, rectanglestyle, bag] {Evaluate the window function $W_{2DF}(\omega_{t},\boldsymbol{R}_{e}(T),\boldsymbol{P}_{e}(T); e(T))$ \\ for the desired population times \textit{T}};
\node (convergence) [below = 1cm and 3cm of evaluation, diamondstyle, bag] {Convergence of $S_{2DF}(\omega_{\tau},T,\omega_{t})$ achieved?};
\node (endsimulation) [below = 1cm and 3cm of convergence, rectanglestyle_round] {End simulation};
\draw [arrow_style] (initialization) --  (propagation);
\draw [arrow_style] (propagation) --  (evaluation);
\draw [arrow_style] (evaluation) --  (convergence);
\draw [arrow_style] (convergence) -- ++ (6,0) |- node[pos=0.25,left,anchor=west] {No} (initialization);
\draw [arrow_style] (convergence) -- node[anchor=east] {Yes} (endsimulation);
\end{tikzpicture}
\end{adjustbox}
    \caption{Flowchart depicting the simulation protocol for obtaining 2D-FLEX spectra as proposed in this work.}
    \label{fig:flowchart}
\end{figure}
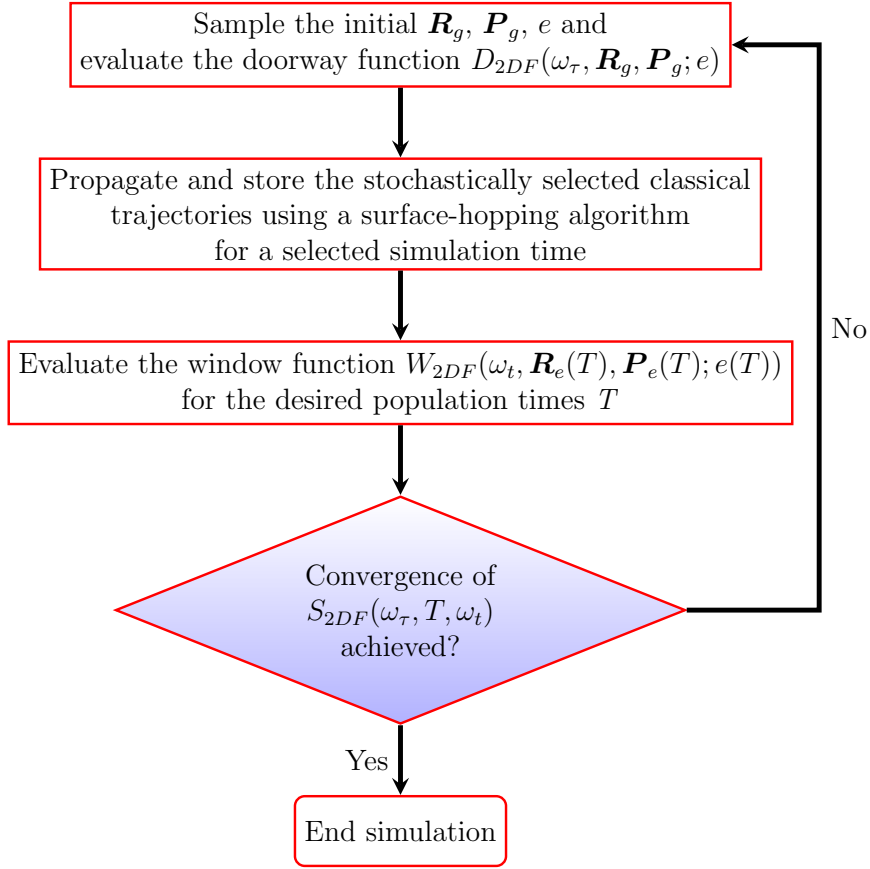


\begin{figure}
    \centering
    \includegraphics[width=1.0\textwidth]{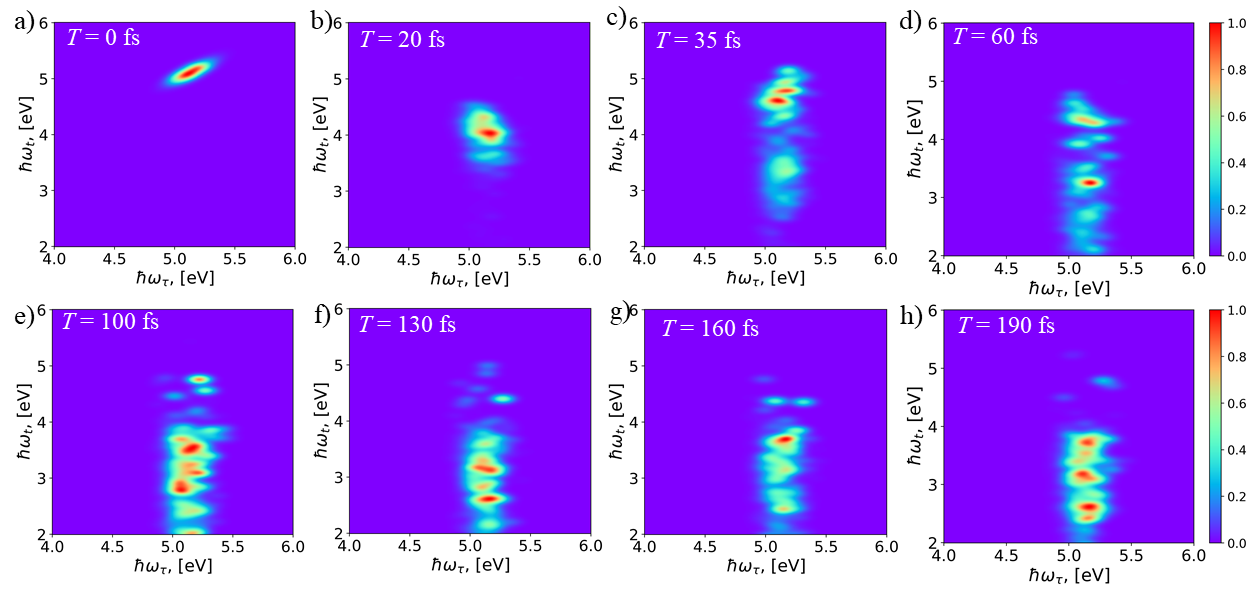}
    \caption{2D-FLEX spectra $S_{\mathrm{2DF}}(\omega_{\tau},T,\omega_{t})$ of pyrazine as a function of the excitation frequency $\omega_{\tau}$ and the detection frequency  $\omega_{\mathrm{t}}$  at population times \textit{T} = 0 (a), 20 (b), 35 (c), 60 (d), 100 (e), 130 (f), 160 (g), 190~fs (h) for $\tau_{pu}$ = 5~fs and $\tau_{t}$ = 30~fs. For each $T$, the intensities of the spectra are scaled for better visibility.}
    \label{fig:2d_evolution_tau5}
\end{figure}

Figure~\ref{fig:2d_evolution_tau5} depicts 2D-FLEX spectra $S_{2DF}(\omega_{\tau},T,\omega_{t})$ of pyrazine as a function of the excitation frequency $\omega_{\tau}$ and the fluorescence frequency  $\omega_t$  at population times \textit{T} = 0 (a), 20 (b), 35 (c), 60 (d), 100 (e), 130 (f), 160 (g) and 190~fs (h). The pulse durations for the pump and gate pulses are set to $\tau_{pu}=5$~fs (with $E_{pu}(\omega)$ plotted in red in Figure \ref{pulses}) and $\tau_t=30$~fs, respectively, which corresponds to the achievable time resolution in fluorescence up-conversion experiments. \cite{Joo05,Joo08} 2D-FLEX spectra excited with longer pump pulses $\tau_{pu} = 15$~fs exhibit similar structure, but the peaks are narrower along $\omega_{\tau}$ (see Figure~S1 in Supporting Information). The spectra in Figure~\ref{fig:2d_evolution_tau5} illustrate the ability of 2D-FLEX to directly monitor the downhill (conical-intersection-mediated) population transfer. Namely, the $\omega_{\tau}$-position of all peaks reveals the initially excited bright $B_{2u}$ state, while the $\omega_{t}$-positions of the peaks follow the $B_{2u} \rightarrow B_{3u}$ population transfer and subsequent energy redistribution in the coupled $B_{3u}/A_u$ states.

At \textit{T} = 0~fs (Figure~\ref{fig:2d_evolution_tau5}a), a single peak around $\omega_{\tau}$ = $\omega_t$ = 5.2~eV   corresponds to the emission from the initially populated  bright $B_{2u}$ state. The peak is elongated along the main diagonal, which signifies a substantial inhomogeneous broadening.\cite{ZS23} As \textit{T} increases, the wavepacket moves down to $\omega_{t} \approx 4$~eV at $T=20$~fs (Figure~\ref{fig:2d_evolution_tau5}b), monitoring the conical-intersection-driven $B_{2u} \rightarrow B_{3u}$ internal conversion process. The spectrum splits into two branches at \textit{T} = 35~fs with stronger peaks in the upper branch and weaker peaks in the lower branch (Figure~\ref{fig:2d_evolution_tau5}c). At \textit{T} = 60~fs (Figure~\ref{fig:2d_evolution_tau5}d), the separation between those two branches disappears, and the spectrum exhibits pronounced doublets at $\omega_t \approx$ 4.4 and 3.2~eV as well as numerous weaker peaks in the entire frequency range $2 \,\, \text{eV} < \omega_t < 4.5 \,\, \text{eV}$. These peaks reveal the lower vibronic levels of the $B_{2u}$ state and higher vibronic levels of the $B_{3u}/A_u$ states. For $T\geq 100$~fs, the spectrum splits again into upper ($\omega_{t} > 4.3$~eV) and lower ($\omega_{t} < 4$~eV) branches. At \textit{T} = 100~fs, the upper branch is dominated by a relatively pronounced peak around $\omega_{t} = 4.7$~eV (Figure~\ref{fig:2d_evolution_tau5}e), and the lower branch is composed of several well-resolved peaks in the frequency range of $\omega_{t}$ = 2.0 - 4.0~eV which mirror the wavepacket motion in the coupled $B_{3u}/A_u$ states. The intensities of peaks in both upper and lower branches decrease with $T$, indicating the irreversible $B_{2u} \rightarrow B_{3u}/A_u$ population transfer (see Figures~\ref{fig:2d_evolution_tau5}e-h). From \textit{T} = 0~fs to \textit{T} = 190~fs, the maximum intensities of the 2D-FLEX spectra decrease by a factor of about 36. The complete time evolution of 2D-FLEX spectra with a time step of 5~fs for \textit{T} are shown in Figures S2-S6 in the Supporting Information.   

\begin{figure}
    \centering
    \includegraphics[width=1.0\textwidth]{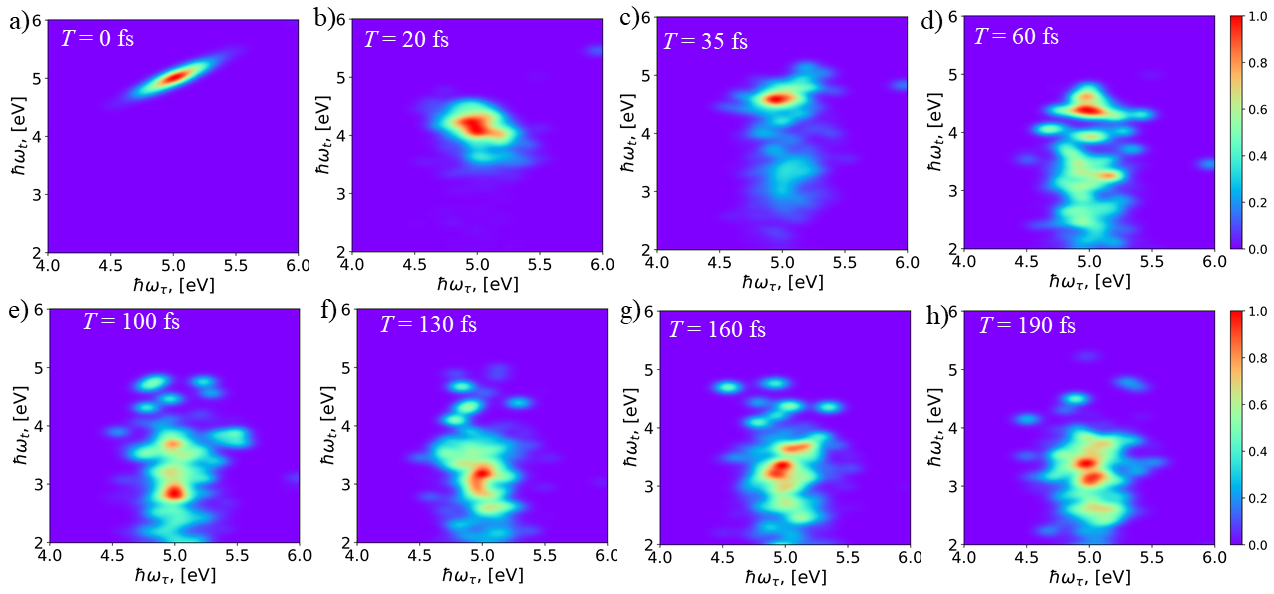}
    \caption{2D-ES SE spectra $S_{SE}(\omega_{\tau},T,\omega_{\mathrm{t}})$ of pyrazine as a function of the excitation frequency     $\omega_{\tau}$ and the detection frequency  $\omega_{\mathrm{t}}$  at population times \textit{T} = 0 (a), 20 (b), 35 (c), 60 (d), 100 (e), 130 (f), 160 (g), 190~fs (h) for $\tau_{pu}$ = $\tau_{pr}$ = 0.1~fs. For each $T$, the intensities of the spectra are scaled for better visibility.}
    \label{fig:2d_evolution_conv_tau01}
\end{figure}

The 2D-ES SE spectra simulated with pulse durations of $\tau_{pu}=\tau_{pr}=$ 5~fs at different population times \textit{T}  monitor exclusively the time evolution of a single peak around $\omega_{\tau}=\omega_t=$ 5.2~eV, while the peaks for lower  $\omega_t$ completely vanish (see Figure S7 in the Supporting Information). This is because the laser pulse with duration of $\tau=$ 5~fs only covers the $B_{2u}$ band of absorption spectrum of pyrazine, but does not overlap with the $B_{3u}/A_u$ band (see the energy profile of the laser pulse in Figure \ref{pulses}). To cover the entire frequency domain  of the absorption spectrum of pyrazine, one needs to use nowadays unavailable attosecond pulses. The 2D-ES SE spectra simulated with such pulses  ($\tau_{pu}$ = $\tau_{pr}$ = 0.1~fs, almost impulsive limit) at population times $T=0$ (a), 20 (b), 35 (c), 60 (d), 100 (e), 130 (f), 160 (g), 190~fs (h)  are displayed in Figure~\ref{fig:2d_evolution_conv_tau01}. Comparison of  Figures~\ref{fig:2d_evolution_tau5} and \ref{fig:2d_evolution_conv_tau01} shows that positions of the peaks  of 2D-ES SE and 2D-FLEX spectra coincide, as expected. Furthermore, intensities of the peaks in 2D-ES SE spectra decrease by a factor of 44 from $T=0$ to $T=190$ fs, which is only slightly larger than a factor of 36 for 2D-FLEX spectra.
However, different detection schemes affect significantly the spectral resolution and peak shapes of the two signals. While 2D-FLEX spectra in Figure~\ref{fig:2d_evolution_tau5} exhibit distinct peaks along $\omega_t$, the peaks of 2D-ES SE spectra in Figure~\ref{fig:2d_evolution_conv_tau01} are much broader. The reason for this difference is as follows. The widths of peaks of 2D-ES SE spectra in Figure~\ref{fig:2d_evolution_conv_tau01} are primarily determined by the optical dephasing parameter of 0.05 eV and are almost independent of the spectral widths of the laser pulses (impulsive limit). The widths and shapes of peaks of 2D-FLEX spectra in  Figure~\ref{fig:2d_evolution_tau5} are predominantly determined by the spectral widths of the pump and gate pulses. 

Comparison of simulated 2D-FLEX and 2D-ES SE spectra demonstrates that different detection schemes can affect the signal dramatically. Detection of 2D-ES SE spectra is  determined by the carrier frequencies and bandwidths of the pulses (cf. Ref.~\cite{Meech17}). As a result, obtaining 2D-ES SE spectra of pyrazine in the entire energy domain of $2 \, \text{eV} \le \omega_t \le 5.2 \, \text{eV}$  requires attosecond pulses. On the other hand, detection of 2D-FLEX spectra in a broad range of wavelengths of spontaneously emitted photons allows fine-tuning of the simultaneous time- and frequency-resolution,\cite{muk4,WhatPhotons} which in turn yields a highly informative and detailed view of the intramolecular energy transfer and redistribution. 
Hence, 2D-FLEX can become a method of choice for studying complex  nonadiabatic dynamics which could only be monitored by 2D-ES with still unavailable  attosecond pulses.

In summary, we develop an on-the-fly SH protocol for the simulation of 2D-FLEX spectra of polyatomic chromophores. If necessary, the developed simulation routine can be further enhanced by employing the machine-learning technique following the recipes of Ref. \cite{pios_pyrazine2024}. The simulated 2D-FLEX spectra  of pyrazine  demonstrate that three facets ensure  high predictive power and information content of this novel spectroscopic technique. (i) Similar to 2D-ES, 2D-FLEX spectra  monitor wavepacket motions of molecular systems excited at frequency $\omega_{\tau}$ and probed at frequency $\omega_{t}$. 
(ii) In contrast to 2D-ES, 2D-FLEX spectra contain exclusively the SE contribution. (iii) 2D-FLEX spectra provide sufficient temporal and spectral resolution by employing the experimentally available pump ($\tau_{pu} \sim 5$ fs) and gate ($\tau_{t}  \sim 30$ fs) pulses, as well as cover the spectral range of several eV along $\omega_t$. Separately, the facets (i)-(iii) are inherent in a number of spectroscopic techniques. For example, time-resolved fluorescence detects solely SE contribution with sufficient temporal ($T$) and frequency  ($\omega_{t}$) resolution,\cite{Joo05,Joo08} and UV/Vis 2D-ES SE spectra permit broadband detection.\cite{Riedle13,Lienau24} It is the combination of all three facets that makes 2D-FLEX spectroscopy a highly promising tool for scrutinizing nonadiabatic dynamics in a number of molecular systems. It is expected that the present work will promote practical  implementation of 2D-FLEX spectroscopy and facilitate interpretation of the forthcoming 2D-FLEX experiments.

\section{Supporting Information}
Doorway-Window functions in 2D-FLEX, Computational details, 2D-FLEX spectra for $\tau_{\mathrm{pu}}$ = 15~fs, the complete time evolution of 2D-FLEX spectra for $\tau_{\mathrm{pu}}$ = 5~fs, 2D-ES SE spectra for $\tau_{pu}=\tau_{pr}$=5 fs.

\begin{acknowledgement}
S.V.P. and L.P.C. acknowledge support from the starting grant of research center of new materials computing of Zhejiang Lab (No.~3700-32601). M.F.G. acknowledges support from the National Natural Science Foundation of China (No.~22373028). 
L.V. acknowledges postdoctoral fellowship of Hangzhou Dianzi University.
J.H. acknowledges support from the Deutsche Forschungsgemeinschaft (DFG, German Research Foundation), Project No. 514636421.
\end{acknowledgement}
\section{Data availability statement}
The data is available from the authors upon reasonable request.
\section{Code availability statement}
The code for the generation of 2D-FLEX spectra from \textit{ab initio} data is available at \url{https://github.com/psebastianzjl/2d_flex_spectra_generation}.
\section{Competing interests}
The authors declare that they have no competing interests.
%
%

\bibliography{main}


\end{document}